\documentclass[conference]{IEEEtran}

\usepackage{atbegshi}
\AtBeginDocument{\AtBeginShipoutNext{\AtBeginShipoutDiscard}}
\usepackage{textpos}
\usepackage[]{algorithm2e}

\usepackage{mathrsfs}
\usepackage{multirow}
\usepackage{enumerate}
\usepackage{comment}
\usepackage{tabularx}
\usepackage{amsmath}
\usepackage{breqn}
\usepackage{paralist}
\usepackage{float}
\usepackage{amsfonts}
\usepackage{lettrine}
\usepackage{balance}
\usepackage{mathtools}
\usepackage{amssymb}
\usepackage[font=small,skip=0pt]{caption}
\usepackage{amssymb}
\usepackage{float}
\usepackage[dvipsnames]{xcolor}
\usepackage{graphicx}
\usepackage{gensymb}
\usepackage{lipsum} 
\usepackage{array}
\usepackage{color,soul}
\usepackage[short]{optidef}
\IEEEoverridecommandlockouts
\usepackage[utf8]{inputenc}
\usepackage[noadjust]{cite}
\usepackage[caption=false,font=normalsize,labelfont=sf,textfont=sf]{subfig}
\usepackage{textcomp}
\usepackage{soul}
\usepackage[export]{adjustbox}
\usepackage{url}
\usepackage[usestackEOL]{stackengine}
\usepackage{comment}

\usepackage{makecell}

\DeclareCaptionLabelSeparator{periodspace}{.\quad}
\begin{document}

\title{Leveraging Data-Driven Models for Accurate Analysis of Grid-Tied Smart Inverters Dynamics}

\author{\IEEEauthorblockN{Sunil~Subedi}
\IEEEauthorblockA{Electrical Engineering and Computer Science \\
South Dakota State University, Brooking, SD, USA\\
Oak Ridge National Laboratory, Oak Ridge, TN, USA\\
Email: sunil.subedi@jacks.sdstate.edu}

\and
\IEEEauthorblockN{Nischal~Guruwacharya}
\IEEEauthorblockA{Electrical Engineering and Computer Science \\
South Dakota State University, Brooking, SD, USA\\
Email: nischal.guruwacharya@jacks.sdstate.edu}

\and
\IEEEauthorblockN{Bidur~Poudel}
\IEEEauthorblockA{Electrical Engineering and Computer Science\\
South Dakota State University, Brooking, SD, USA\\
Email: bidur.poudel@jacks.sdstate.edu}

\and
\IEEEauthorblockN{Jesus D.~Vasquez-Plaza}
\IEEEauthorblockA{Electrical \& Computer Engineering Department,\\  University of Puerto Rico\\
Mayagüez Campus, Mayagüez, Puerto Rico, USA\\
Email: jesus.vasquez@upr.edu}

\and
\IEEEauthorblockN{Fabio~Andrade}
\IEEEauthorblockA{Electrical \& Computer Engineering Department,\\  University of Puerto Rico\\
Mayagüez Campus, Mayagüez, Puerto Rico, USA\\
Email: fabio.andrade@upr.edu}

\and
\IEEEauthorblockN{Robert~Fourney}
\IEEEauthorblockA{Electrical Engineering and Computer Science \\
South Dakota State University, Brooking, SD, USA\\
Email: robert.fourney@jacks.sdstate.edu}

\and
\IEEEauthorblockN{Hossein~Moradi~Rekabdarkolaee}
\IEEEauthorblockA{Mathematics and Statistics \\
South Dakota State University, Brooking, SD, USA\\
Email: hossein.moradirekabdarkolaee@sdstate.edu}

\and
\IEEEauthorblockN{Timothy~M.~Hansen}
\IEEEauthorblockA{Electrical Engineering and Computer Science\\
South Dakota State University, Brooking, SD, USA\\
Email: timothy.hansen@sdstate.edu}

\and
\IEEEauthorblockN{Reinaldo~Tonkoski}
\IEEEauthorblockA{Electrical \& Computer Engineering \\
University of Maine, Orono, Maine, USA \\
Electric Power Transmission \& Distribution \\
Technical University of Munich, Munich, DE\\
Email: tonkoski@ieee.org}
}

\thanks{
\begin{minipage}{\textwidth}
This manuscript has been authored by UT-Battelle, LLC, under contract DE-AC05-00OR22725 with the US Department of Energy (DOE). The US government retains and the publisher, by accepting the article for publication, acknowledges that the US government retains a nonexclusive, paid-up, irrevocable, worldwide license to publish or reproduce the published form of this manuscript, or allow others to do so, for US government purposes. DOE will provide public access to these results of federally sponsored research in accordance with the DOE Public Access Plan (http://energy.gov/downloads/doe-public-access-plan).

This work is supported by the U.S. Department of Energy Office of Science, Office of Electricity Microgrid R\&D Program, and Office of Energy Efficiency and Renewable Energy Solar Energy Technology Office under the EPSCoR grant number DE-SC0020281.
\end{minipage}}

\maketitle

\begin{minipage}{\textwidth}
\vspace{10 em}
\centering
    Datasets are available at https://data.mendeley.com/datasets/82k8x5tpkn/2~\cite{Subedi_2023}.
\end{minipage}

\clearpage

\begin{abstract}
Integrating power electronic converters (PECs) and distributed energy resources (DERs) into power systems has led to a dynamic system and the need for accurate simulation to understand the impact of converter domination on the power grid. The fast switching and stochastic dynamics of inverter-based resources (IBRs) in converter-dominated power systems (CDPS) require accurate models for analysis. When it comes to developing models for real-world systems, a few different approaches can be taken. Although different modeling methodologies exist, black-box or data-driven system identification (SysId) techniques create PEC models from experimental data without knowledge of the internal system physics. This approach involves a system identification process that includes selecting a model class, estimating model parameters, and validating the model. While there are different linear and nonlinear model structures and estimation algorithms available, it's important to be creative and understand the physical system to create a data-driven model that works well for the intended application, whether that be simulation, prediction, control, or fault detection. This report provides information on collecting datasets from commercial off-the-shelf (COTS) inverters and detailed simulation models.

\end{abstract}

\maketitle

\section{Introduction}

The increasing integration of power electronic converters (PECs) and distributed energy resources (DERs) in power systems has led to a shift from a historically passive to a more dynamic system~\cite{ieee2021,Chinmay_access}. As rooftop and utility-scale solar photovoltaics, energy storage, electric vehicles, and controllable loads with converters become integrated with the power grid, any changes in their operation will have a significant impact on the entire grid when they dominate it (known as a converter-dominated power system (CDPS))~\cite{sunil_nrel}. It is crucial for engineers to simulate the system to understand the effects of converter domination on the power grid. PEC models are crucial to determining the stability impacts of inverter-based resources (IBRs), developing control strategies to increase the integration of renewable energy, and exploring other opportunities and challenges in renewable energy and energy storage.

These IBRs feature faster switching processes and more stochastic dynamics than traditional power systems. However, the impact of these dynamics has been largely overlooked in the past due to the low fraction of IBRs and the converters' passive role in voltage and frequency management, as well as the dominance of well-defined models for traditional power systems driven by large synchronous generators~\cite{kundur1994power}. In the context of CDPS, it is essential to investigate these dynamics as they have implications for controller design, optimization, supervisory actions, fault detection, component diagnostics, and instability prediction of complex systems. Reliable models are needed to study the dynamics of CDPS with IBR generation~\cite{Chinmay_access,sunil_access}.

According to the IEEE 1547-2018 standard for DERs and other grid codes, PECs with various grid support functions (GSFs) provide voltage and frequency ancillary services. However, depending on the manufacturer, these services can be implemented differently even under similar operating conditions~\cite{ieee1547}. While these smart PECs with GSFs are meant to facilitate the integration of more IBRs and regulate grid voltage and frequency, their specific design and control actions are often proprietary and unknown. This can lead to errors in power system modeling, simulation, and PEC dynamics, resulting in inaccurate results and analysis. Furthermore, the dynamics of PECs can vary depending on the activation of states and the GSF standards used~\cite{pll_sunil_2022}. Detailed switching models based on the most accurate PEC models from manufacturers, if available, may not be suitable for real-time control as they are computationally intractable. Therefore, there is a need for data-driven PEC models with GSFs in CDPS that do not rely on knowledge of the control structure or physical topology~\cite{sunil_access}.

Several modeling methodologies in the literature aim to represent the dynamics of PECs while reducing computational costs accurately. Switched models are the most thorough approach for capturing non-linear dynamics, but they become computationally demanding as the number of PECs increases~\cite{switchmodel2014}. Averaged linear models reduce computational costs by ignoring switching dynamics, but are less accurate in representing all of the process dynamics~\cite{average_2001}. Dynamic phasor models are more accurate than average models, but are technically more complex~\cite{Raja2013}. Black-box or data-driven system identification (SysId) techniques, on the other hand, create PEC models from experimental data with little or no knowledge of the physics of the internal system (internal parameters, topology/control structure)~\cite{pico2018,nischal2020,Valdivia}. This work uses this approach to create models for PECs in CDPS while providing multiple ancillary services.

To help to improve in CDPS analysis, this report provides detailed information on datasets collected from commercial off-the-shelf (COTS) inverters and simulated studies. This report is a companion document to the data hosted on the Mendeley data port. The remainder of this report is organized as follows. Section~\ref{Dynamic_modeling} begins with introducing the concepts of black-box modeling representing the dynamics, probing signal, and the concept of the partitioned modeling. Section~\ref{casestudy} describes the simulation setup for data collection followed by the data collection and details of the data. Finally, the paper is concluded in Section~\ref{conclusion}.

\section{Dynamic Modeling of PECs with GSFs}\label{Dynamic_modeling}

The authors used the SysId process to capture the nonlinear dynamics of a photovoltaic (PV) inverter operating in different GSFs. This involves explaining the various probing signals used and the modeling approach to model the system. A detailed flow chart is presented in ~\cite{nischal2022speedam_pv_inv_modeling} to outline the steps in obtaining the reduced-ordered model of the inverter is presented in~\cite{nischal2022speedam_pv_inv_modeling}.

\subsection{SysId of PECs Dynamics with GSFs}

The article focuses on utilizing SysId to obtain a mathematical representation of various real and simulation inverters. The 3-phase Fronius Symo Inverter (FSI) and 1-phase SMA real inverters located in the SDSU microgrid laboratory and the simulation-based 1-phase PV PEC are subjected to testing. However, the methodology can be applied to any vendor's PEC inverter model.

To apply SysId accurately, it is necessary to choose the correct (1) excitation signals, (2) model input, (3) model architecture, (4) system dynamics representation, (5) model order, (6) model structure and complexity, (7) model parameter and (8) model validation. The complexity and need for prior knowledge often diminish from step (1) to step (7). In general, steps (1) through (7) are carried out using measurement data referred to as ``training data," while step (8) is usually carried out using measurement data referred to as ``validation data". 

\begin{figure}[h!]
    \centering
    \includegraphics[width=\columnwidth]{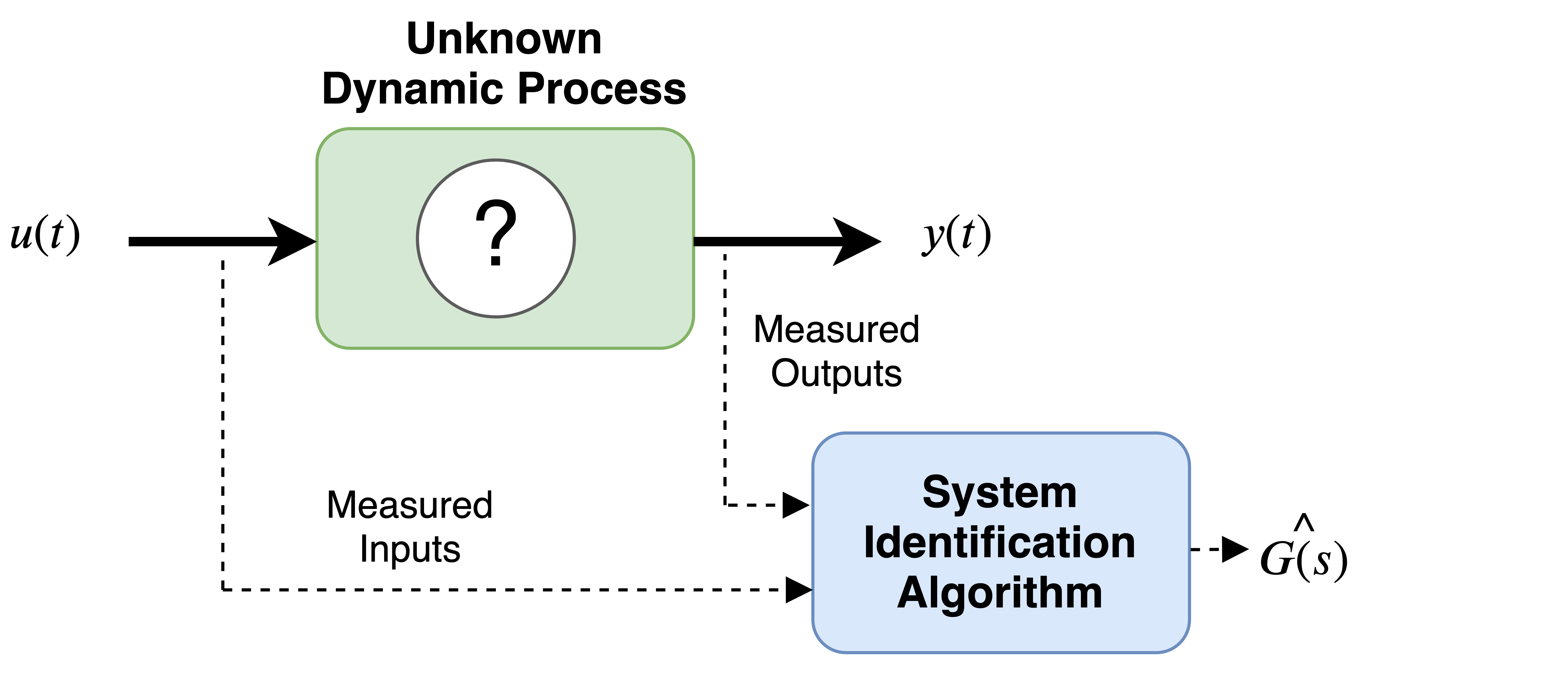}
    \caption{The graphic demonstrates the fundamental concept of SysId. The SysId approach uses input and output data to identify an unknown dynamic process. Performance is assessed using statistical measures.}
    \label{fig:identification}
\end{figure}

The training set is then passed through a SysId algorithm to estimate the system parameters and obtain a reduced-ordered model. The accuracy of the obtained model is validated using the testing dataset, and the goodness-of-fit is calculated based on the normalized root-mean-square error (NRMSE). The article highlights the benefits of using SysId as a data-driven approach to modeling PECs in CDPS without requiring knowledge of the control structure or parameters. Fig.~\ref{fig:identification} depicts the essential notion of a SysId process.

\subsection{Probing Signals}

Proper probing signal design is crucial to learning about the system's behavior in the parameterization of the data-driven model~\cite{Manisha2020, nelles_2021}. Constraints from power systems and SysId theory must be followed, and rectangular/square signals are ideal for estimating time constant and emphasizing frequency range~\cite{pierre2010signals}. The design of the probing signal should consider noise, signal amplitude, and the desired frequency range. The logarithmic, square chirp, which involves the sweep of the logarithmic frequency of a square wave, is the most accurate in a study of four different signals to extract dynamics of COTS inverters~\cite{nischal2022speedam_pv_inv_modeling}. It will be used as the probing signal for perturbing the system.

\subsection{Metric for quantitative assessment}\label{Metrics}

After the data was collected and the model was fitted, it was required to validate the estimated model by evaluating its statistical performance. A collection of models with varying numbers of poles and zeros can fit the data based on the acquired input-output data. A goodness of fit measure based on NRMSE is calculated as~\cite{MATLABtoolbox_fit}:

\begin{equation}
    \text{NRMSE} = \frac{\left \| y(t) - \hat{y}(t) \right \|_2}{\left \| y(t) - {\;}{\bar{y}(t)} \right \|_2}
\end{equation}

\begin{equation}
    \label{eqn:fitpercentage}
    \textit{fitpercent} = 100\times\left(1 - \text{NRMSE}\right)\%
\end{equation}

where $||.||_2$ indicates the 2-norm of the vector, $y(t)$ denotes data from the detailed PEC simulation model, $\hat{y}(t)$ denotes data from the aggregated reduced-ordered simulation model, and $\bar{y}(t)$ indicates the channel-wise mean of the measured output data. The goodness-of-fit in~(\ref{eqn:fitpercentage}) indicates how well the response of the derived models fits the estimation data. Higher the goodness-of-fit, the better the model. 

\subsection{Partitioned Modeling Approach}

The dynamic behavior of PECs while operating in various GSFs modes is complex due to nonlinearities, making it difficult to model the entire operating region with a single linearized model~\cite{Sunil}. To overcome this, the operating regions are divided into multiple linear regions based on the magnitude of voltage and frequency defined in IEEE~\cite{ieee1547}. Each region (R) is further divided into smaller ranges (r), and the SysId algorithm is applied to calculate a reduced-ordered linear dynamic model for each range. Combining these reduced models allows dynamic for the entire region can be obtained~\cite{Valdivia}. The regions and ranges of three different GSFs, i.e., Volt-VAr, Volt-Watt, and Freq-Watt characteristics curves, are detailed in~\cite{gsf_nis}. 

\section{Case Study}\label{casestudy}

\subsection{3-phase Fronius Symo Inverter with Volt-VAr}\label{3-phase}

\subsubsection{Simulation Setup}

To determine the reduced-ordered dynamic model of a commercial Fronius Symo Inverter, a power hardware-in-the-loop (PHIL) setup was used. The equipment under test was a 10 kW three-phase FSI operating in Volt-VAr mode. The PHIL setup consisted of a real-time digital simulator (RTDS) and a perturbation and measurement unit to emulate the grid voltage. The voltage amplitude at the point-of-common-coupling (PCC) was perturbed, and the current injected by the FSI into the grid was recorded over its normal operating range of 0.88 to 1.1 p.u. A PV system powered the DC side of the FSI, and a resistive load was connected to the PCC to consume the generated power and protect against reverse power flow.  FSI and load settings are detailed in~\cite{nischal2022speedam_pv_inv_modeling}.

\begin{figure}[!ht]
    \centering
    \includegraphics[scale = 0.95]{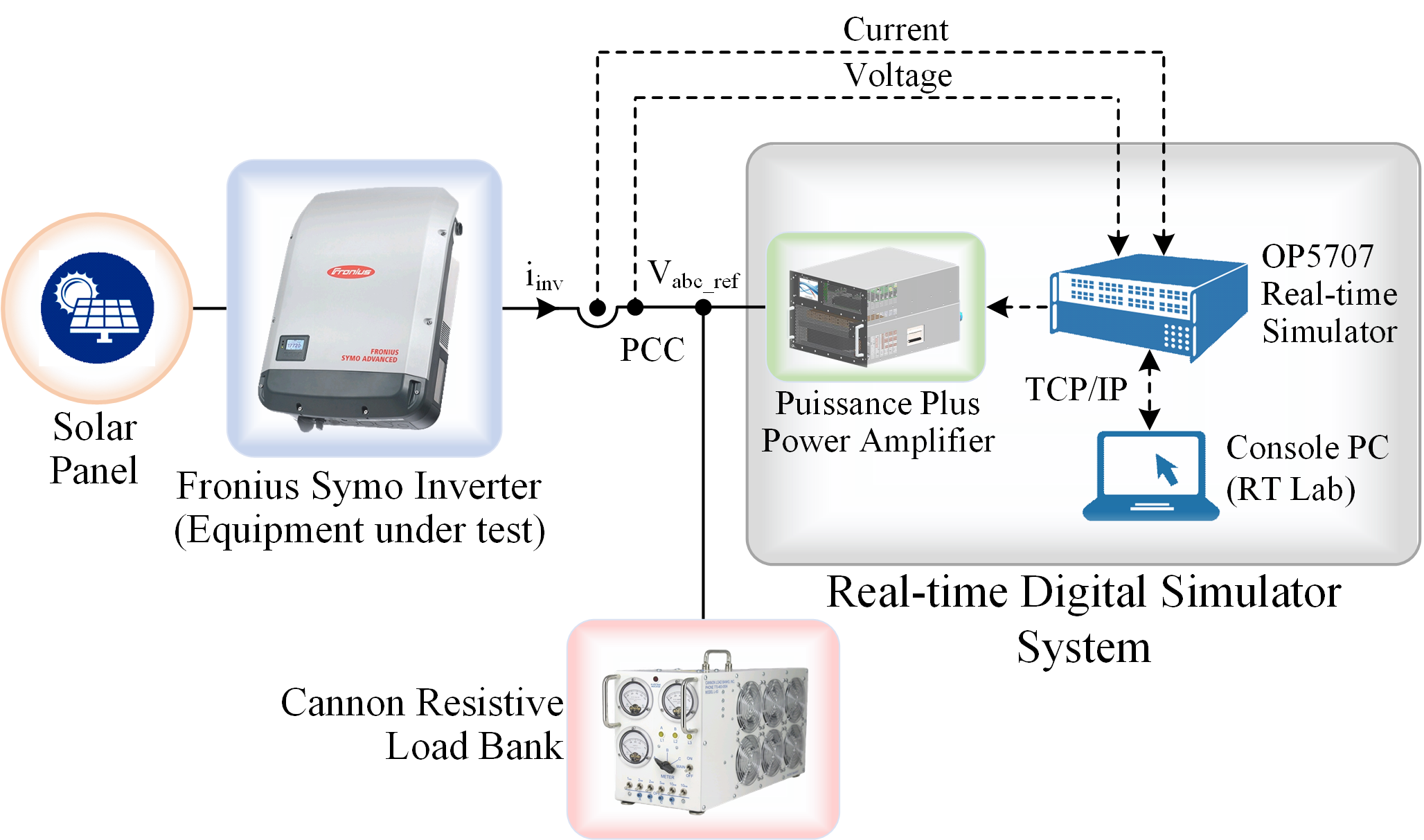}
    \vspace{5pt}
    \caption{Experimental setup to determine the dynamic reduced-ordered model of single-phase Fronius Symo inverter operated in Volt-VAr mode. The inverter is probed through a Puissance Plus Power Amplifier unit controlled through an Opal-RTDS.}
    \label{fig:experimental_setup}
\end{figure}

\begin{table}[!ht]
    \centering
    \caption{Load and Fronius Symo Inverter Parameters~\cite{ieee1547}.}
    \label{tab:inverter_parameter_3phasae}
    \begin{tabular}{|l|l|}
    \hline
    \textbf{Parameter}          & \textbf{Value} \\\hline\hline
    Resistive load & 1 kW \\\hline
    PV power available & 85 W\\\hline
    $Q_1$, $Q_2$, $Q_3$, and $Q_4$ & 3.3, 0, 0, and -3.3 kVA resp.\\\hline
    $V_L$, $V_1$, $V_2$, $V_3$, $V_4$, and $V_H$ & 0.88, 0.92, 0.98, 1.02, 1.08,\\
     &  and 1.1 p.u. resp.\\\hline
    \end{tabular}
\end{table}

\subsubsection{Data Extraction}

To determine the reduced-order dynamic model, the voltage amplitude in the PCC and the current injected by the FSI was recorded for the complete normal Volt-VAr operating range (0.88 p.u. to 1.1 p.u.). The voltage amplitude was increased gradually over a 15 s period, and four different types of signals (Sine, Square, Sine-Chirp, and Square-Chirp) were used to perturb the voltage. The frequencies of the signals were selected based on the settling time response parameters of the inverter, with the Sine and Square waves having a frequency of 1 Hz and the Sine-Chirp and Square-Chirp signals having a frequency range of 1 Hz to 32 Hz.

The data collected were filtered, divided into regions, and divided into training and validation sets for cross-validation. Poles and zeros were swept from a second-order model to a fifth-order model, and the most accurate model was selected based on the lowest Akaike final prediction error (AFPE). The final reduced-ordered model for each range was derived by aggregating all of the selected transfer functions (TFs), allowing for replicating the inverter's dynamics for the entire operating region.
\vspace{10 pt}

\subsubsection{Data Explanation and Results}

The PHIL test's voltage parameters in regions R$_1$ to R$_5$ were divided into ranges (3, 7, 3, 5, and 2) such that each ranges are equally spaced. To test the models, the data in the middle range, i.e., $(r_{12}, r_{24}, r_{32}, r_{43})$ was used to obtain the TF for R$_1$ to R$_4$, while the data in the range 1, i.e., $(r_{51})$ was used for R5. However, data from any range will work, The second-order TF was used to represent the overall dynamics of the FSI due to its accuracy and computational efficiency. R$_3$ was not considered in the analysis because the Volt-VAr mode is deactivated. Different probing signals were used to determine the TFs of FSI in Volt-VAr mode, resulting in 16 TFs. The Sq-Chirp signal performed the best in all regions when comparing the TFs obtained from various probing signals. The high fit percentage obtained with the Sq-Chirp signal is because it contains higher frequency components that can capture the dynamics of FSI.

The collected data set is available in Mendeley source in~\cite{Subedi_2023}.
The collected data for each range can be found in the MAT file named ``$volt\_var\_R.mat$" in the data port. The data were sampled at a rate of $100~\micro$s, and each range was excited for a maximum of 15 s. The system identification algorithm for the TFs is provided as ``$volt\_varl\_R.m$" on the data port. The first 110 s of the data were removed for analysis because the inverter takes that time to synchronize with the grid. The split-up data was also subjected to a move-median filter with a window size 200, and initial transients of 0.5 s were removed. The input and output mean values were removed before applying the system identification algorithm. The data was divided into a training dataset (70\%) and a testing dataset (30\%) to generate the TFs. After the TFs were generated, the mean value of the current was added to the estimated current to obtain the actual estimated current from the TF model. Finally, to test the TFs, the testing signal was used, and \textit{fitpercent} for each TF was calculated.

The file has a total of 9 columns;

\begin{enumerate}
    \item time (s)
    \item reference voltage (from the simulation) (volt). It needs to be converted into p.u. ($(120\times sqrt(2))$)
    \item voltage amplitude at PCC (volt). Need to converter into p.u. by dividing ($(120\times sqrt(2))$).
    \item direct current (I$_d$) of an inverter (Amp)
    \item quadrature current (I$_q$) of an inverter (Amp)
    \item I$_q$ current obtained from derived TF developed with the square wave as input (Amp)
    \item I$_q$ current obtained from derived TF developed with the logarithmic square chirp as input (Amp)
    \item I$_q$ current obtained from derived TF with sinusoidal as input (Amp)
    \item I$_q$ current obtained from derived TF with logarithmn sinusoidal chirp as input (Amp)
\end{enumerate}

Using this data, TF is obtained.

After running the code, the obtained transfer function for various regions is shown below:\\

 \noindent R$_1$ $\rightarrow$ $\frac{-511.5s+9.886e04}{s^2+121.8s+7098}$\\
 
\noindent  R$_2$ $\rightarrow$ $\frac{-1667s+9.49e04}{s^2+31.84s+456.3}$ \\

 \noindent R$_4$ $\rightarrow$ $\frac{-139.5s+402.4}{s^2+110.2s+1785}$ \\

\noindent R$_4$ $\rightarrow$ $\frac{-2017s+1.054e05}{s^2+32.55s+484.9}$ \\

 \noindent R$_5$ $\rightarrow$ $\frac{548.7s- 1.193e05}{s^2+187.8s+1.187e04}$ 

\subsection{SMA Inverter with Volt-VAr and Volt/Freq-Watt}

\subsubsection{Simulation Setup}

A PHIL setup tests a commercial two-phase 5 kW SMA inverter under the IEEE 1547-2018 standard voltage and frequency settings. The experiment setup includes a PV system, the SMA inverter, a Puissance Plus Power Amplifier, RTDS, a Resistive Load Bank, and a Console PC. The RTDS and power amplifier perturb the voltage and frequency at the PCC, while the load bank utilizes the PV power and protects against reverse power flow. The dynamics of the inverter are captured by recording the perturbed voltage and frequency at PCC and the current injected by the SMA inverter. The inverter operates in Volt-VAr and either in Volt-Watt or Freq-Watt mode, which is selected based on the IEEE standard. The logged data is passed through the Opal-RT system and analyzed using the SysId algorithm in MATLAB/Simulink.

\begin{figure}[!ht]
    \centering
    \includegraphics[width=\columnwidth]{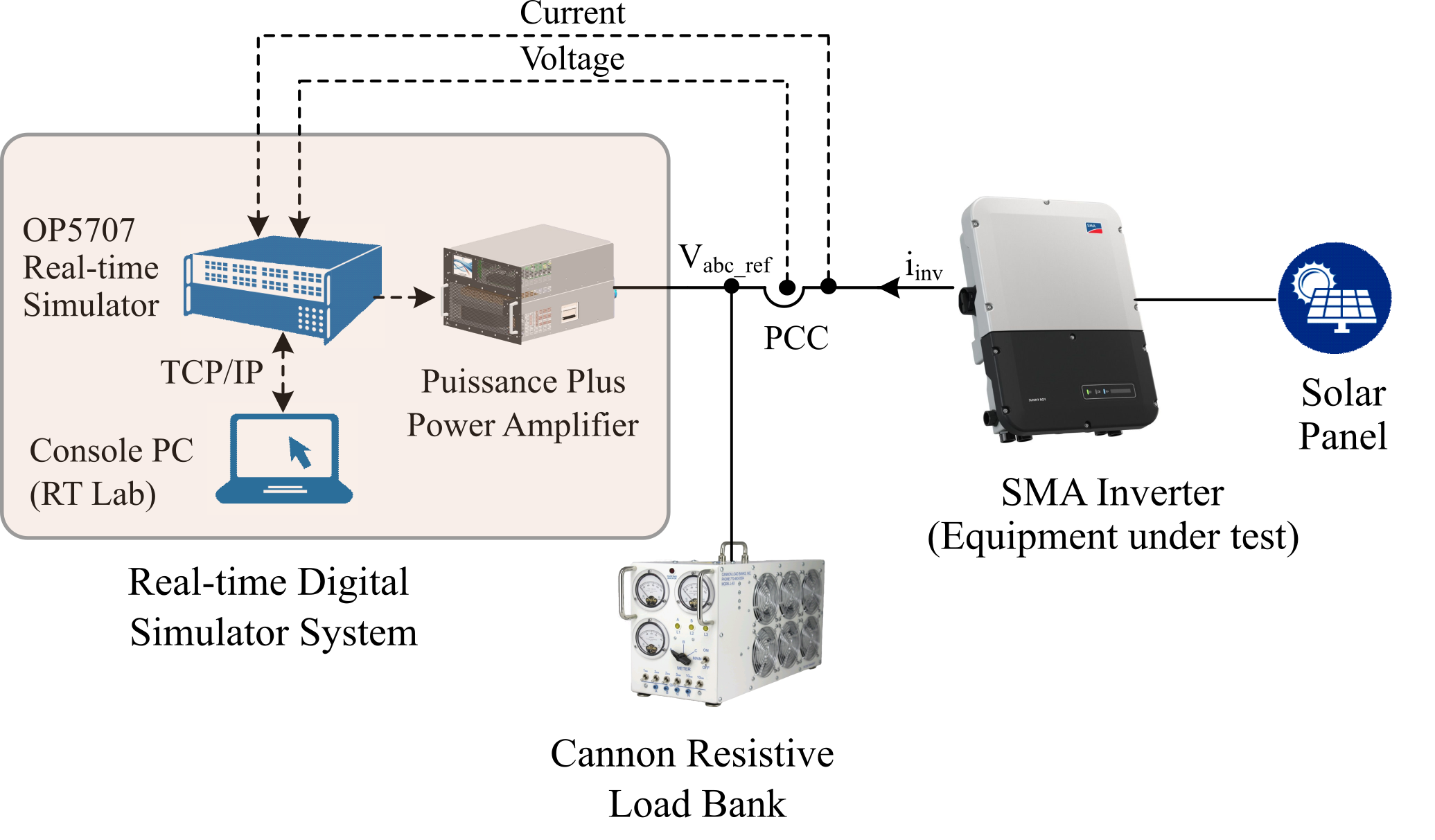}
    \vspace{1pt}
    \caption{Experimental setup to determine the dynamic reduced-ordered model of single-phase SMA inverter operated in Volt-Watt and Freq-Watt modes. The inverter is probed through a Puissance Plus Power Amplifier unit controlled through an Opal-RTDS.}
    \label{fig:Experimental Setup}
\end{figure}

\begin{table}[!ht]
    \centering
    \caption{Load and SMA Inverter Parameters.}
    \label{tab:inverter_parameter_twophase}
\begin{tabular}{cc}
\hline
\multicolumn{1}{c}{\textbf{Parameter}}      & \textbf{Value}                        \\ \hline \hline
\multicolumn{1}{c}{Resistive load}          & 1 kW                                  \\ \hline
\multicolumn{1}{c}{SMA Rating}              & 5 kW                                  \\ \hline
\multicolumn{2}{c}{\textbf{Volt-Watt Setting}}                                               \\ \hline\hline
\multicolumn{1}{c}{$P_1$
 and $P_2$} & 3, 0 and 0 kW resp.                  \\ \hline
\multicolumn{1}{l}{$V_1$, $V_2$ and $V_H$} & 1.045, 1.085 and 1.095 p.u. resp. \\ \hline
\multicolumn{2}{c}{\textbf{Freq-Watt Setting}}                                               \\ \hline \hline
\multicolumn{1}{c}{$P_1$ and $P_2$} & 3, 0 and 0 kW resp.                  \\ \hline
\multicolumn{1}{c}{$F_1$, $F_2$, and $F_H$} & 1.0083, 1.033, and 1.037 p.u. resp. \\ \hline
\end{tabular}
\end{table}

\subsubsection{Data Extraction}

When testing the inverter under the IEEE 1547-2018 standard, the setup includes a PV system, SMA inverter, power amplifier, RTDS, resistive load bank, and console PC. The system changes the voltage and frequency amplitude of a logarithmic, square chirp probing signal to determine the dynamics of the inverter operating in Volt-VAr and Volt-Watt or Freq-Watt mode. The frequencies of the signals were selected based on the settling time response parameters of the inverter, with the logarithmic, square chirp signals having a frequency range of 0.5 Hz to 5 Hz. In Volt-VAr, the voltage amplitude changes from 0.88 to 1.095 p.u. In Volt-Watt mode, the voltage amplitude changes from 1 p.u. to 1.095 p.u. with a step of 0.005 p.u. and the frequency amplitude is kept constant at 1 p.u. In Freq-Watt mode, the frequency amplitude changes from 1 p.u. to 1.0370 p.u. with a step of 0.0015 p.u. The voltage amplitude is kept constant at 1 p.u. A band-pass filter is used to remove noise.\\

\subsubsection{Data Explanation and Results}

\noindent{\textbf{Volt-VAr:}}

Solar Irradiance data were collected for three periods of the day, i.e., morning, mid-day, and evening to analyze the effect of the varying irradiance in the data-driven modeling of the COTS inverter. The data were sampled at a rate of $100\micro$s, and each range was excited for a maximum of 15 s. In Volt-VAr mode, a similar approach is utilized as explained in ~\ref{3-phase}. The collected data for each range can be found in the data port in the MAT file named \small{``$Sq\_Chirp\_VV\_SMA.mat$"}~\cite{Subedi_2023}. The data were sampled at a rate of $100\micro$s, and each range was excited for a maximum of 15 s. First,  The system identification algorithm used to find the TFs is provided as \small{``$Load\_to\_Simulink.m$"} in the data port. The file has a total of 9 columns with similar input and output as explained in~\ref{3-phase}.\\

\noindent{\textbf{Volt-Watt:}}

These three data sets for Volt-Watt (available in the data port) were collected during the morning, mid-day, and evening time to analyze the impact of varying irradiance in the dynamic modeling of the inverter.
\begin{itemize}
    \item $Sq\_Chirp\_SMA\_VW\_Morning.mat$
    \item $Sq\_Chirp\_SMA\_VW\_Midday.mat$ 
    \item $Sq\_Chirp\_SMA\_VW\_Evening.mat$ 
\end{itemize}

  In each file, the required input and output are the actual reference voltage and current injected by the inverter. The system identification algorithm used to find the TF is provided as \small{``$Load\_to\_Simulink.m$"} in the data port.\\

 In each .mat file, the required input and output are: 
\begin{enumerate}
    \item (first column) time (s)
    \item (second column) reference voltage (from the simulation) whose unit is in volt. It needs to be converted into p.u. [input]
    \item (fourth column) Id current injected by inverter (Amp) [response or output]
\end{enumerate}

To model the inverter dynamics in Volt-Watt and Freq-Watt mode, the mid-day dataset ($r_{13}$, $r_{24}$, and $r_{31}$) was used, and the partitioned modeling approach was employed. In Volt-Watt mode, the regions R$_1$, R$_2$, and R$_3$ were divided into 5, 7, and 2 smaller ranges, respectively, and the TFs of the respective regions were obtained using the mid-range data. The analysis of region R$_3$ was not considered in this case because the Volt-Watt mode of the inverter was turned off in that region, and the inverter could not inject or absorb any active power. During the PHIL test condition in Freq-Watt mode, the regions R$_1$, R$_2$, and R$_3$ are divided into 7, 10, and 2 smaller ranges based on the frequency settings.

\noindent{\textbf{Freq-Watt:}}

The three data sets (available in the data port) were collected during the morning, mid-day, and evening time to analyze the impact of varying irradiance in the dynamic modeling of the inverter.
\begin{itemize}
    \item $Sq\_Chirp\_SMA\_FW\_Morning.mat$
    \item $Sq\_Chirp\_SMA\_FW\_Midday.mat$ 
    \item $Sq\_Chirp\_SMA\_FW\_Evening.mat$ 
\end{itemize}
The system identification algorithm in the data port provides the TF as \small{``$Load\_to\_Simulink.m$"}. \\

In each .mat file, the required input and output are: 
\begin{enumerate}
    \item (first column) time (s)
    \item (second column) reference frequency (from the simulation) whose unit is in Hz. It needs to be converted into p.u. [input]
    \item (fourth column) Id current injected by inverter (Amp) [response or output]
\end{enumerate}

In the case of Freq-Watt mode, the analysis of regions R$_1$ and R$_3$ is discarded as the SMA inverter will prevent absorbing/injecting any active power in those regions. The morning and evening datasets were used with the estimated mid-day TF to analyze the effect of varying irradiance on inverter dynamics. However, the output response showed a need for a DC-gain adjustment in the mid-day TF model. To further validate this requirement, the TF models were derived for the morning and evening datasets using the same procedure as the mid-day dataset to obtain the DC-gain adjustment value.

\subsection{1-phase and 3-phase PV Inverter with Volt-VAr}

\subsubsection{Simulation Setup}

\begin{figure}[!ht]
    \centering
    \includegraphics[width=\columnwidth]{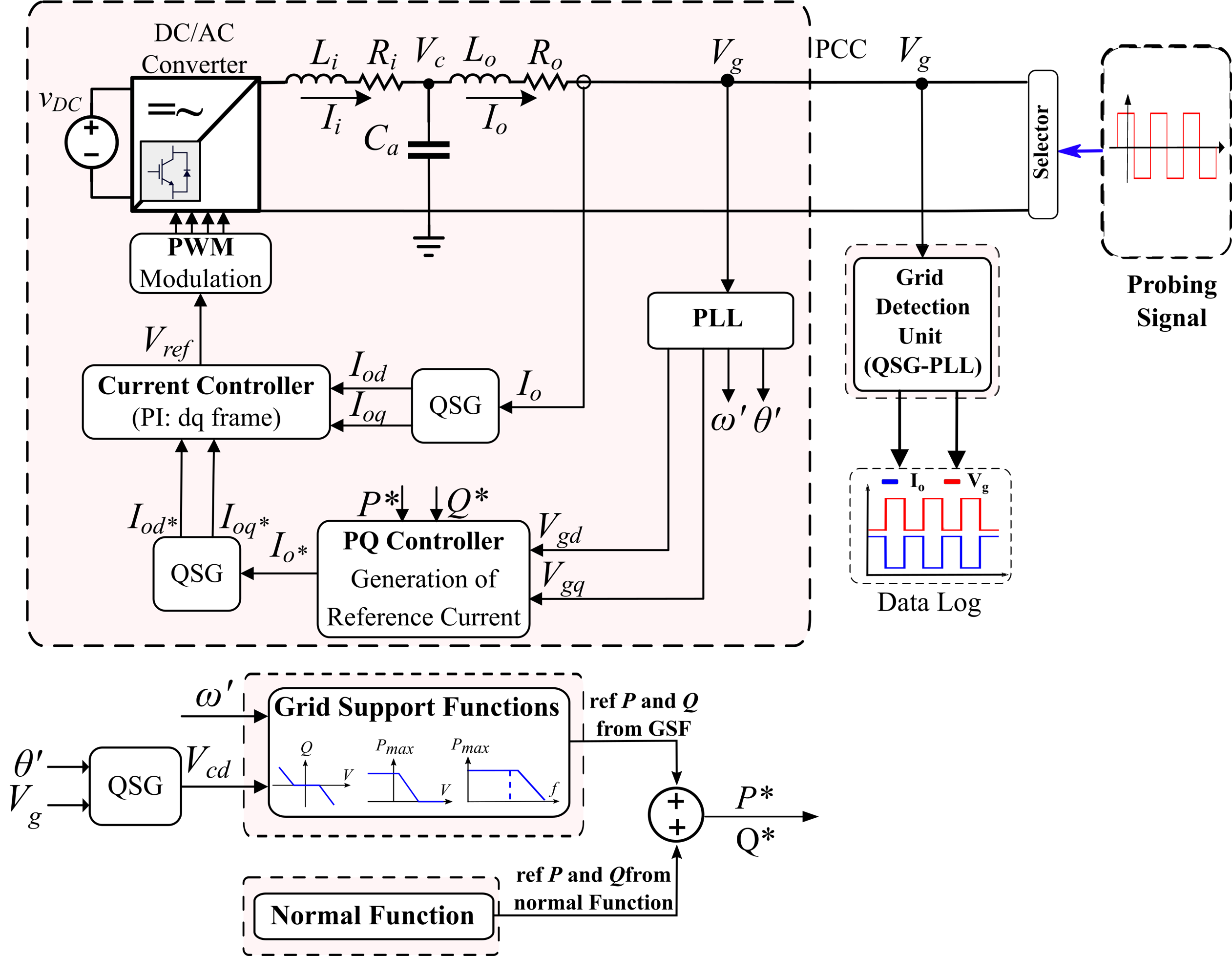}
    \vspace{6 pt}
    \caption{Simulation setup diagram of the grid-connected system with the various components and control loops in a smart inverter with normal and GSFs features.}
    \vspace{6 pt}
    \label{fig:inv_schematic_1p}
\end{figure}

Figure~\ref{fig:inv_schematic_1p} illustrates a simulation setup of a switched single-phase voltage source PEC. The PEC is powered by a continuous DC bus and connected to the primary grid through an LCL filter to reduce switching noise. The variables $I_i$, $V_c$, $I_o$, and $V_g$ represent input current, capacitor voltage, output current, and grid voltage, respectively. A proportional-integral (PI) controller was utilized in the $d$-$q$ architecture to regulate the $I_o$ sent to the primary grid by the VSI. The second-order generalized integral (SOGI) - phase-lock loop (PLL) was selected because of its simplicity of installation and excellent performance~\cite{sogi2013,pll_sunil_2022}. However, detail of PLL impacts on data-driven modeling can be found in~\cite{pll_sunil_2022}. The PLL, also known as SOGI-PLL in the report, is used to measure the magnitudes of the $V_g$ and $I_o$ generated by PECs while following the fundamental frequency ($\omega'$) and phase angle ($\theta'$). Depending on the condition, the PEC can operate in feeding or supporting mode, injecting constant or varying P and Q into the grid. The PQ controller generates reference current sent to the current controller, affecting the grid current's dynamic responsiveness. 

\begin{table}[!ht]
    \centering
    \caption{Load and PEC Parameters}
    \label{tab:inverter_parameter_1phasesim}
\begin{tabular}{cc}
\hline

\multicolumn{1}{c}{\textbf{Parameter}}      & \textbf{Value}                                 
                                             \\ \hline\hline
\multicolumn{2}{c}{\textbf{Transformer}}                                               \\ \hline\hline
\multicolumn{1}{c}{$S$, $V_1$, $V_2$} & 75 kVA, 14.4 kV, 240 V                   \\ \hline
                                             
\multicolumn{2}{c}{\textbf{Inverter}}                                               \\ \hline\hline
\multicolumn{1}{c}{$P$ and Available Power} & 8.4 kW and 6.25 kW                   \\ \hline

\multicolumn{2}{c}{\textbf{Volt-VAr Setting}}  
            \\ \hline   

\multicolumn{1}{c}{$Q_1$, $Q_2$, $Q_3$ and $Q_4$} &  6.25, 0, 0, and -6.25 kW                  \\ \hline
\multicolumn{1}{l}{$V_L$, $V_1$, $V_2$, $V_3$, $V_4$, and $V_H$ } & 0.88, 0.92, 0.98, 1.02, 1.08, 1.10 p.u.\\ \hline
\end{tabular}
\end{table}

\subsubsection{Data Extraction}

Single-phase inverter operating at Volt-VAr mode is subjected to the test.
The voltage at the point of PCC is perturbed within normal operating limits set by IEEE, between 0.88 and 1.1 p.u. A logarithmic, square chirp signal with a voltage amplitude that starts at 0.88 p.u. and increases by 0.001 p.u. every 6 seconds until it reaches 1.1 p.u. is used. Based on the PEC's settling time response, the frequency is swept from 1 to 5 Hz. A step-change of a minimum of 0.01 p.u. was chosen because, after a measurement resolution of 0.01, the measurement noise dominates the input. The detail of the setup, experiment, and method is explained in~\cite{subedi2023automated}.

\subsubsection{Data Explanation and Results}

We update the algorithm used in our previous work, which divides the voltage magnitudes evenly to achieve the recommended technique. Partition modeling with a binary search algorithm is implemented in Python to generate many reduced-order TFs by varying the number of poles and zeros and calculating the fit percent of each model. The best reduced-order TF for each range is selected based on its fit percent, and these selected models also correspond to the models with the highest Adj. $R^2$ and the lowest corrected Akaike Information Criterion (AICc) and Bayesian Information Criterion (BIC). The voltage parameters in regions R$_1$ to R$_5$ were divided into 22 ranges based on a minimum of 1 and a maximum of 22 considering a measurement resolution of 0.01 ranging from 0.88 to 1.1 p.u. obtained from the partitioned modeling algorithm. By varying poles and zeros, we got 22 unique TFs models considering the trade-off between accuracy and computational complexity. We aggregated them to represent the overall dynamics of the FSI. A logarithmic square-chirp signal was used to determine the TFs of FSI in Volt-VAr mode.

The collected data set is in the source under the folder ``Simulation Data-driven Modeling of 1phase smart PV inverter"~\cite{Subedi_2023}.
The collected data for each range can be found in the MAT file named ``$volt\_curr.mat$" in the data port. The data were sampled at a rate of $100\micro$s, and each range was excited for a maximum of 6 s. The chirp frequency was varied from 1--5 Hz. The SysId algorithm implemented in Python is used to find the TFs. The initial 1 s of the data was removed for analysis since the inverter takes that time to synchronize with the grid. Data were divided into training and testing datasets to generate the TFs. After the TFs were generated, the mean value of the current was added to the estimated current to obtain the actual estimated current from the TF model. Finally, the testing signal was used, and \textit{fitpercent} for each TF was calculated. The plot of the input and output is given in ``volt\_curr.py".

In ``$volt\_curr.mat$" file, the required input and output are: 
\begin{enumerate}
    \item (first column) time (s)
    \item (third column) reference voltage whose unit is in p.u.
    \item (fourth column) Iq current injected by inverter (Amp) [response or output]
\end{enumerate}

To validate the derived TFs models, step response was given in each region (i.e. R$_1$ to R$_5$) with the respective range as amplitude (i.e. 0.88 to 0.92 p.u., 0.92 to 0.98 p.u., 0.98 to 1.02 p.u., 1.02 to 1.08 p.u. and 1.08 to 1.1 p.u.). The collected data for each range can be found in the MAT file named ``$step\_r1.mat$", ``$step\_r2.mat$", ``$step\_r3.mat$", ``$step\_r4.mat$", and  ``$step\_r5.mat$" in the data port. The performance can be visualized with ``$Step\_Resonse.ipynb$".

In `each mat file, the required input and output are: 
\begin{enumerate}
    \item (first column) time (s)
    \item (second column) response from the detailed inverter model
    \item (third column) response from the reduced ordered model \end{enumerate}

The collected data set for three-phase PV inverter is in the source under the folder ``Simulation Data-driven Modeling of 3phase smart PV inverter".
The collected data for each range can be found in the MAT file named ``$volt\_curr2.mat$" in the data port. The data were sampled at a rate of $100\micro$s, and each range was excited for a maximum of 10 s. The chirp frequency was varied from 1--5 Hz. The SysId algorithm implemented in Python is used to find the TFs. The initial 1 s of the data was removed for analysis since the inverter takes that time to synchronize with the grid. Data were divided into training and testing datasets to generate the TFs. After the TFs were generated, the mean value of the current was added to the estimated current to obtain the actual estimated current from the TF model. Finally, the testing signal was used, and \textit{fitpercent} for each TF was calculated.

In ``$volt\_curr2.mat$" file, the required input and output are: 
\begin{enumerate}
    \item (first column) time (s)
    \item (first column) grid voltage whose unit is in p.u.
    \item (fourth column) Iq current injected by inverter (Amp) [response or output]
    \item (fifth column) reference input signal in p.u.
    \item (sixth and seventh column) active and reactive power output respectively
    \item (rest 6 columns) voltage and current in abc frame
\end{enumerate}

\subsection{Aggregated Dynamics of Low Voltage Distribution Network}

\subsubsection{Simulation Setup of Low Voltage Distribution Network (LVDN)}

\begin{figure*}[!ht]
   \centering
    \includegraphics[width=0.9\textwidth]{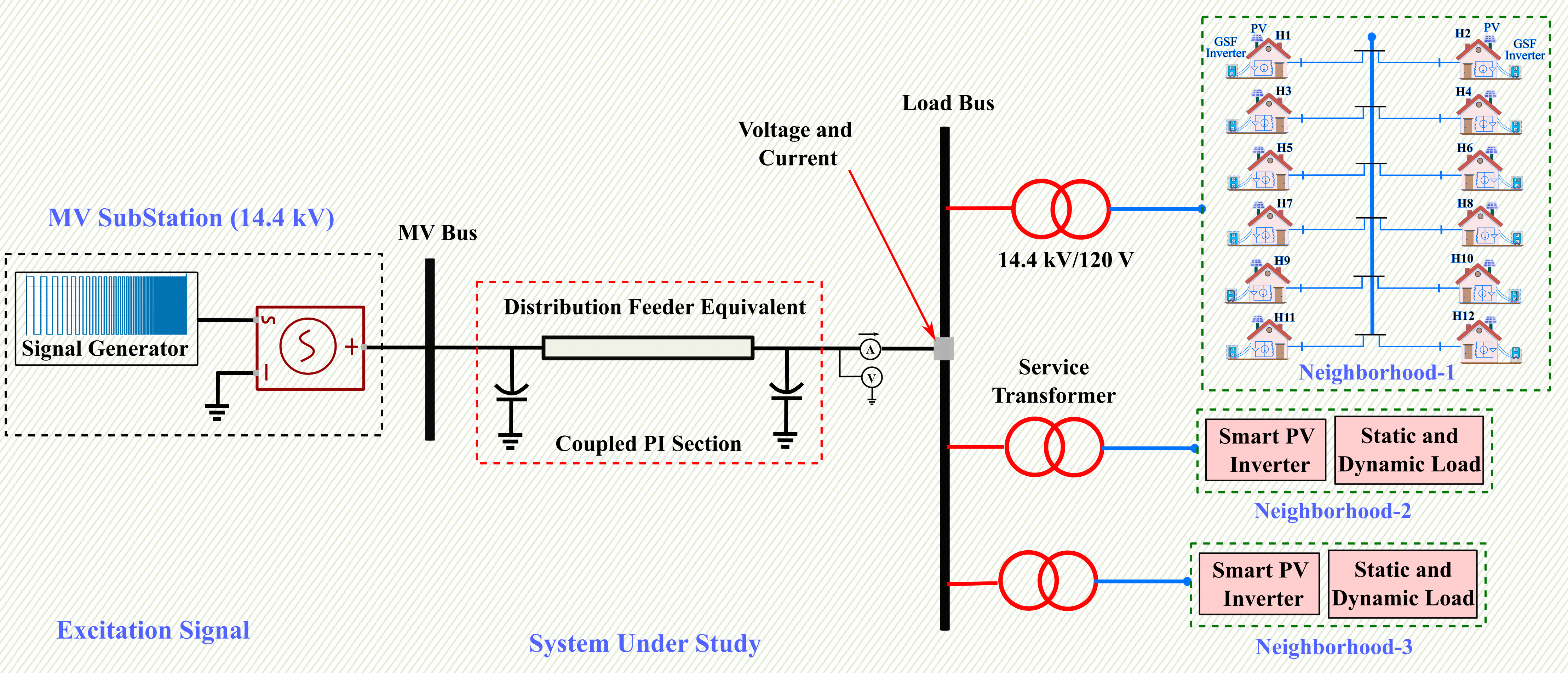}
        \caption{The medium voltage (MV)-feeder line connected to the single/three-phase LVDN neighborhood system (three subcommunities in each phase) with advanced DER inverter and CMLD.}
        \label{fig:Simulation_setup}
\end{figure*}

This section describes the developed DN benchmark with 36 houses, utilizing smart PECs and composite load (CMLD) based on standard feeder characteristics from~\cite{DRT2012}. The LVDN model connects the MV-feeder line, serving as the distribution feeder, to the LVDN system via a load bus (Fig.\ref{fig:Simulation_setup}). The overhead community is situated at the end of a 5km, 14.4kV distribution feeder with 336 kcmil aluminum-stranded conductors. After 5km, the feeder splits into three single-phase laterals for the community and subcommunities. The system includes 36 houses, grouped into subcommunities of 12 homes each. These houses have smart PV inverters (8.4~kW peak capacity, 4.375 kW net power) and 4.375 kW CMLD (ZIP + IM). Similarly, a second scenario is created to test the applicability of the proposed partitioned modeling approach by varying the system (single to three phases) and the PEC controller (PI to Linear Quadratic Integrator (LQI)) and to make a fair comparison with the aggregated state-of-the-art DER\_A model with CMLD.

Subcommunities connect to the load bus through 75kVA, single/three-phase, 14.4kV–240/120 V service transformers with 1.02 p.u. split-phase supply for a maximum 5\% voltage drop. The NS90 feeder type consists of two live wires twisted around a grounded neutral cable (120m length, 100m feeder backbone). Individual houses connect with two wires and a steel grounded neutral cable (20~m drop).
\begin{table}[!ht]

\centering
\caption{Inverter parameter and setting and Probing Signal Design Parameter.}
\label{tab:inverter_parameter_signal}
\begin{tabular}{ll}
\hline
\multicolumn{2}{c}{\textbf{Inverter Parameter}}                                         \\ \hline \hline
Rated Capacity         & \multicolumn{1}{c}{6.25 kW}              \\ \hline
Net-Power & \multicolumn{1}{c}{4.375kW} \\ \hline
$L_i$, $R_i$, $L_o$, $R_o$ & \multicolumn{1}{c}{0.1$\Omega$, 0.0018$H$, 0.1$\Omega$, 0.0018$H$} \\ \hline
$C_a$ & \multicolumn{1}{c}{8.8$\micro F$} \\ 
\hline
\multicolumn{2}{c}{\textbf{Current PI-2 Controller Parameter}}                                         \\ \hline \hline
K$_P$ & \multicolumn{1}{c}{4.5}              \\ \hline
K$_I$ & \multicolumn{1}{c}{60} \\ 

\hline
\multicolumn{2}{c}{\textbf{Volt-VAr Settling}}                                         \\ \hline \hline
$Q_1$, $Q_2$, $Q_3$, $Q_4$         & \multicolumn{1}{c}{4.375, 0, 0, -4.375 kW}              \\ \hline
$V_L, V_1, V_2, V_3, V_4, V_H$ & \multicolumn{1}{c}{0.88, 0.92, 0.98, 1.02, 1.08, 1.1 p.u.} \\ \hline
\multicolumn{2}{c}{\textbf{Probing Signal Parameter}}                                  \\ \hline \hline 
Amplitude (A)                     & \multicolumn{1}{c}{0.8884 - 1.0884 p.u.}                   \\ \hline
Frequency ($f_0$ to $f_1$)                     & \multicolumn{1}{c}{1 to 5 Hz}                                                \\ \hline
Chirp Sweep Time ($T$)              & \multicolumn{1}{c}{5 s}                                                   \\ \hline
Rate of Increase in Frequency $(\lambda$)  & \multicolumn{1}{c}{1\%}                                                   \\ \hline
\end{tabular}

\end{table}

\subsubsection{Simulation Setup of DER\_A with CMLD}

\begin{figure}[!ht]
   \centering
    \includegraphics{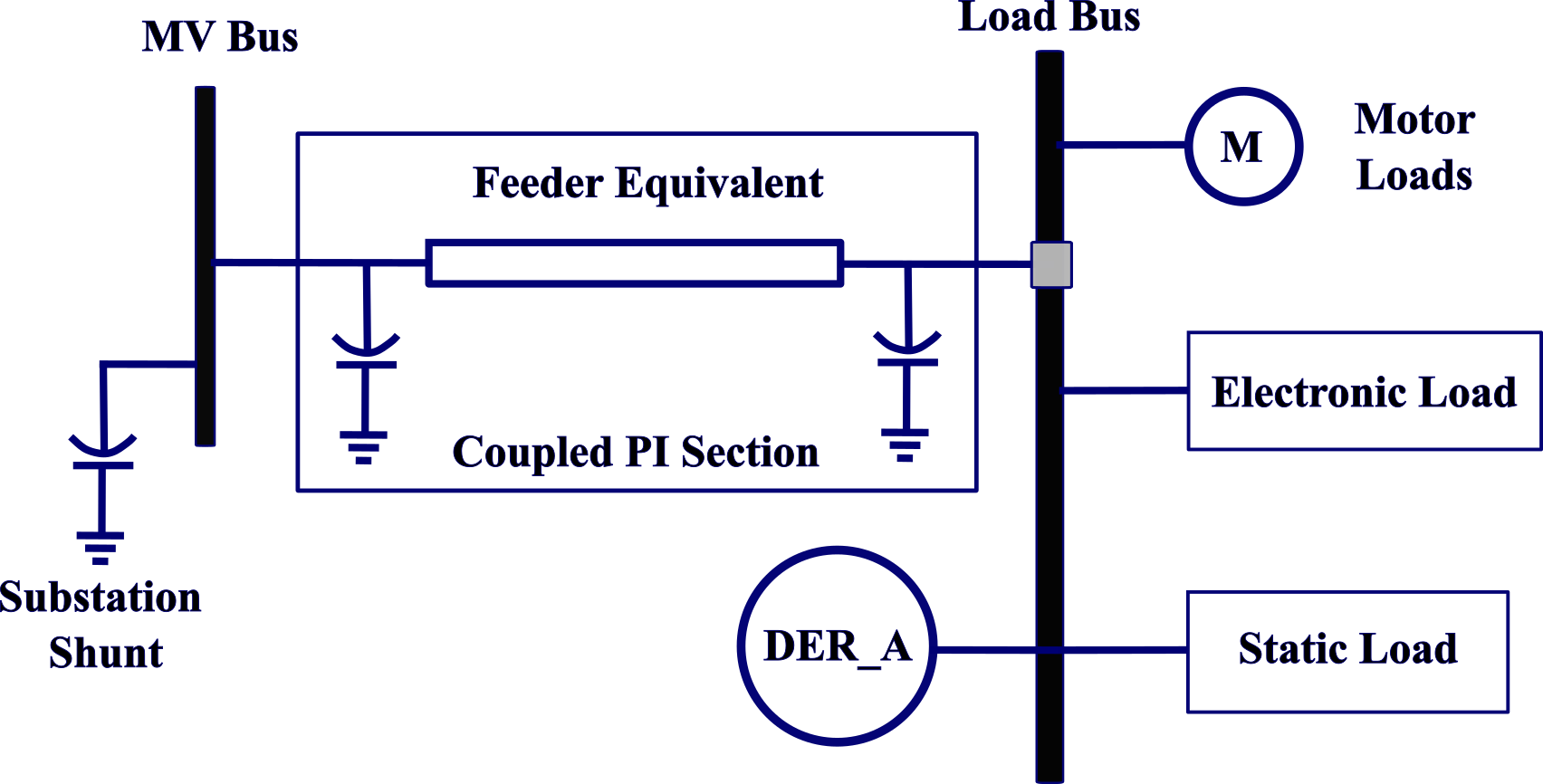}
        \caption{The schematic of CMLD with DER attached.}
        \label{fig:CMLD}
\end{figure}

The schematic diagram of the CMLD is shown in Fig.~\ref{fig:CMLD}, which highlights the combination of ZIP and IM as a prime example of a DER\_A with CMLD.

\begin{figure}[!ht]
    \centering
    \includegraphics[trim={0.2cm 0.2cm 0.06cm 0.2cm},clip,width=\columnwidth]{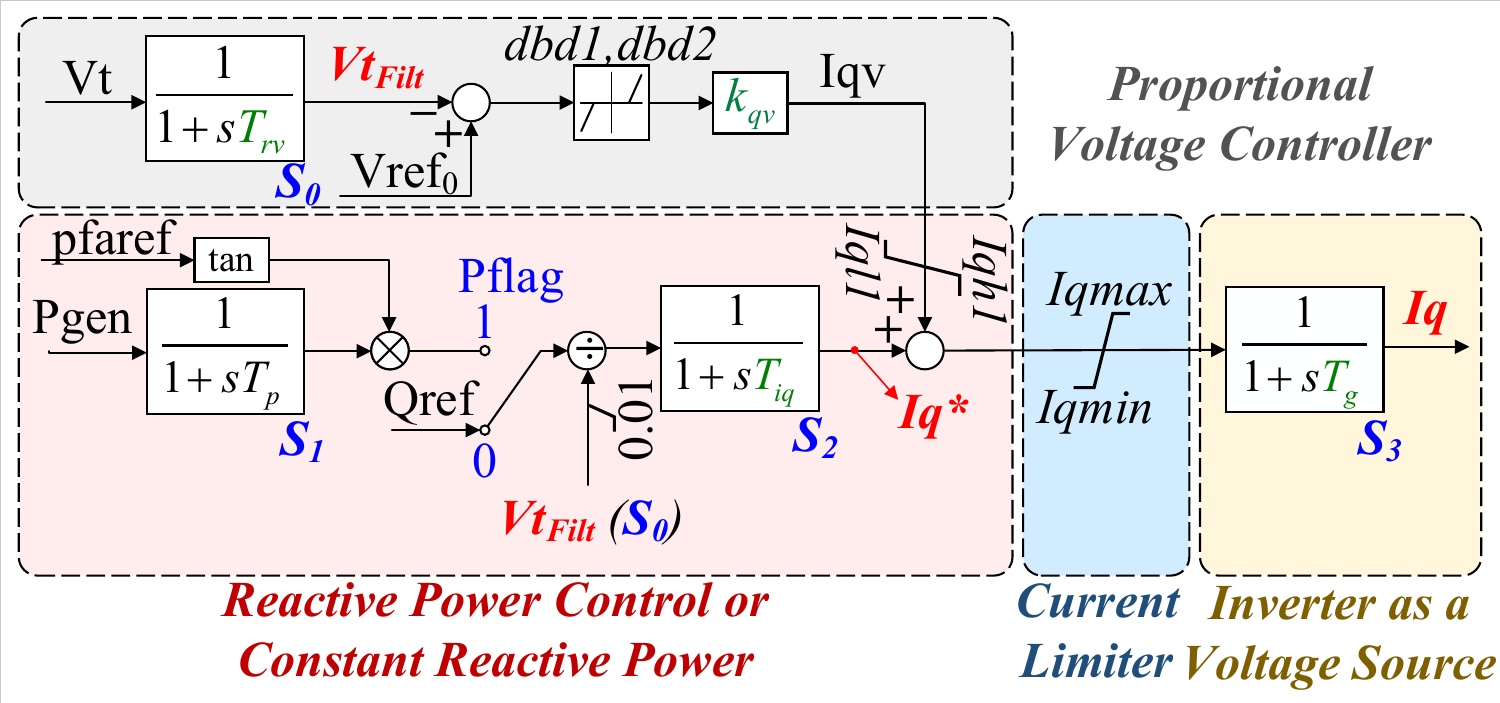}
    \caption{Reactive Power Side (RPS) of Aggregated Model DER\_A}
    \label{fig:DER A_Qside}
\end{figure}

Only the reactive power side (RPS) of the DER A model is used since the LVDN only considers the Volt-Var support function and the RPS is the section that contemplates this function, the rest of the DER\_A model is not considered in this dataset.  Fig.~\ref{fig:DER A_Qside} shows the RPS of the DER\_A model, where $Vt_{Filt}(S_0)$ is the filtered voltage at the model's terminals, $Iq^\ast(S_2)$ is the reactive current provided by the reactive power control, and $Iq(S_3)$ is the total reactive current injected/absorbed. These three are the state variables of the RPS of the DER\_A model. Pflag=1 was used, for that reason state $S_1$ is not considered in this data-set \cite{Jesus_DER_A}.

\subsubsection{Data Extraction}
The signal generator generates the probing signal and excites at the substation.
The voltage and current in the load bus at the point of common coupling are recorded, where the input is the bus voltage. At the same time, the output is the total current, i.e., either injected or absorbed from or by all the neighborhoods connected to the bus. The recorded voltage and current represents the aggregated dynamics of the multiple DERs (PV), PECs, and the CMLD. 

\subsubsection{Data Explanation and Results}

The collected data set is in the source under the folder ``Simulation\_Aggregated\_Data-driven Mofrling of LVDN and DERA\_CMLD"~\cite{Subedi_2023}.
The collected data for each range for single-phase LVDN can be found in the MAT file named ``$LVDN\_Single\_Phase.mat$" in the data port. The data representing the aggregated dynamic behavior of the multiple DERs, PECs, and the load were sampled at a rate of $100\micro$s, and each range was excited for a maximum of 6 s. Similarly, the collected data for each range for three-phase LVDN can be found in the MAT file named ``$LVDN\_Three\_Phase.mat$" in the data port, and the data collected for DER\_A with CMLD can be found in ``$PQ\_VI\_DERA\_CMLD.mat$." The column names of the respective mat files are given in the dot txt file within the folder. 

In ``$LVDN\_Single\_Phase.mat$" file, the required input and output are: 
\begin{enumerate}
    \item (first column) time (s)
    \item (second column) voltage at the point of common coupling at the load bus whose unit is in Volts. It needs to be converted into p.u. [input]
    \item (fourth column) Iq current injected by single LVDN that have the neighborhoods with multiple DERs (PV), PECs, and CMLD (Amp) [response or output]
\end{enumerate}

In ``$LVDN\_Three\_Phase.mat$" file, the required input and output are: 
\begin{enumerate}
    \item (first column) time (s)
    \item (second column) voltage at the point of common coupling at the load bus whose unit is in Volts. It needs to be converted into p.u. [input]
    \item (fourth column) Iq current injected by three LVDN that have the neighborhoods with multiple DERs (PV), PECs, and CMLD (Amp) [response or output]
\end{enumerate}

In ``$PQ\_VI\_DERA\_CMLD.mat$" file, the required input and output are: 
\begin{enumerate}
    \item (first column) time (s)
    \item (fourth column) voltage at the point of common coupling at the load bus whose unit is in Volts. It needs to be converted into p.u. [input]
    \item (seventh column) Iq current injected by DER\_A with CMLD system (Amp) [response or output]
\end{enumerate}

\section{Conclusions}\label{conclusion}
In conclusion, the collection of datasets from both real and simulated inverters and the developed reduced-ordered transfer function models can provide valuable contributions to power system modeling, analysis, planning, operation, and dispatch. By utilizing these datasets and models, engineers and researchers can gain insights into the behavior and performance of inverters under various conditions, which can help inform the design and operation of power systems. Additionally, these models can aid in developing more efficient and effective control strategies, improving power system stability and reliability. Overall, using collected datasets and derived models is crucial in advancing power system engineering and ensuring power systems' efficient and reliable operation.

\section*{Data Availability}
The data is available at https://data.mendeley.com/datasets/82k8x5tpkn/2.

\bibliographystyle{IEEEtran}
\bibliography{references.bib}

\end{document}